\RequirePackage{fix-cm}
\documentclass[twocolumn]{svjour3}          
\smartqed  
\usepackage{graphicx}

\usepackage[caption=false]{subfig}
\usepackage{graphicx,amssymb,xcolor}
\usepackage{url}
\usepackage{amsmath}
\usepackage{multirow}
\usepackage{wrapfig}

\usepackage{xspace}
\newcommand*{\eg}{e.g.\@\xspace}
\newcommand*{\ie}{i.e.\@\xspace}
\newcommand{\etal}{et al.}

\usepackage{hyperref}
\usepackage{tikz}
\usepackage{scalerel}
\usetikzlibrary{svg.path}
\definecolor{orcidlogocol}{HTML}{A6CE39}
\tikzset{
	orcidlogo/.pic={
		\fill[orcidlogocol] svg{M256,128c0,70.7-57.3,128-128,128C57.3,256,0,198.7,0,128C0,57.3,57.3,0,128,0C198.7,0,256,57.3,256,128z};
		\fill[white] svg{M86.3,186.2H70.9V79.1h15.4v48.4V186.2z}
		svg{M108.9,79.1h41.6c39.6,0,57,28.3,57,53.6c0,27.5-21.5,53.6-56.8,53.6h-41.8V79.1z M124.3,172.4h24.5c34.9,0,42.9-26.5,42.9-39.7c0-21.5-13.7-39.7-43.7-39.7h-23.7V172.4z}
		svg{M88.7,56.8c0,5.5-4.5,10.1-10.1,10.1c-5.6,0-10.1-4.6-10.1-10.1c0-5.6,4.5-10.1,10.1-10.1C84.2,46.7,88.7,51.3,88.7,56.8z};
	}
}

\newcommand\orcidicon[1]{\href{https://orcid.org/#1}{\mbox{\scalerel*{
				\begin{tikzpicture}[yscale=-1,transform shape]
					\pic{orcidlogo};
				\end{tikzpicture}
			}{|}}}}

\usepackage[misc,geometry]{ifsym}
\usepackage{etoolbox}
\newcommand*{\affaddr}[1]{#1} 
\newcommand*{\affmark}[1][*]{\textsuperscript{#1}} 
\journalname{International Journal of Information Security (IJIS)}

\begin{document}\sloppy

\title{A Content-Based Deep Intrusion Detection System
}


\author{Mahdi~Soltani\affmark[1] \and
        Mahdi~Jafari~Siavoshani\affmark[1]\and
        Amir~Hossein~Jahangir\affmark[1] \orcidicon{0000-0000-0000-0000}
}


\institute{Mahdi~Soltani \at
              \email{mahdi@ce.sharif.edu}           
           \and
            Mahdi~Jafari~Siavoshani \at
              \email{mjafari@sharif.edu}
           \and
           \Letter		Amir~Hossein~Jahangir \at
           	  \email{jahangir@sharif.edu} \\\\
\affaddr{\affmark[1]Department of Computer Engineering, Sharif University of Technology, Azadi Ave.
	Tehran 1458889694, Iran}
}

\date{Received: date / Accepted: date}

\maketitle

\begin{abstract}
The growing number of Internet users and the prevalence of web applications make it necessary to deal with very complex software and applications in the network. This results in an increasing number of new vulnerabilities in the systems, and leading to an increase in cyber threats and, in particular, \emph{zero-day} attacks. The cost of generating appropriate signatures for these attacks is a potential motive for using machine learning-based methodologies. Although there are many studies on using learning-based methods for attack detection, they generally use extracted features and overlook raw contents. This approach can lessen the performance of detection systems against content-based attacks like SQL injection, Cross-site Scripting (XSS), and various viruses.

In this work, we propose a framework, called deep intrusion detection (DID) system, that uses the \emph{pure content} of traffic flows in addition to traffic metadata in the learning and detection phases of a passive DNN IDS. To this end, we deploy and evaluate an offline IDS following the framework using LSTM as a deep learning technique. Due to the inherent nature of deep learning, it can process high dimensional data content and, accordingly, discover the sophisticated relations between the auto extracted features of the traffic. To evaluate the proposed DID system, we use the CIC-IDS2017 and CSE-CIC-IDS2018 datasets. The evaluation metrics, such as precision and recall, reach $0.992$ and $0.998$ on CIC-IDS2017, and $0.933$ and $0.923$ on CSE-CIC-IDS2018 respectively, which show the high performance of the proposed DID method.
\end{abstract}

%

\section{Introduction} \label{sec:intro}
We live in the cyber era in which network-based technologies have become omnipresent. Meanwhile, threats and attacks are rapidly growing in the cyberspace. Nowadays, mainly signature-based intrusion detection systems (IDSs) are used to detect these malicious traffic. However, since new vulnerabilities and, consequently, zero-day attacks appear each day, the cost of generating accurate signatures with a low false-positive rate is growing.

The traditional approach to intrusion detection systems is based on detecting some form of a \emph{signature}. A signature is extracted from the known attacks by employing security experts. A signature must completely cover different variants of the attack for which it has been extracted. Also, benign traffic and other types of attacks should not be falsely confused with it. Hence, extracting an accurate signature is a complicated and time-consuming process. By the increasing growth of the Internet's applications and users, more vulnerabilities are expected to appear, which results in emerging more new attacks. Therefore, the signature extraction process becomes a more challenging problem in the coming years.

The learning-based approach is an alternative solution to the signature-based intrusion detection systems. In addition to resolving the signature extraction problem, some learning approaches can also detect zero-day attacks by determining abnormal traffic. 

There are several research studies on the use of machine learning methods to detect intrusions in computer networks. Among them, we can mention pioneers like Bayesian networks \cite{Jemili2007Bayesian}, support vector machine (SVM) \cite{Heba2010Principle}, decision trees \cite{Zhang2008RF}, and the new deep learning techniques (\eg, see \cite{Agarap2018GRU} and \cite{Salama2011Deepbelief}). These studies generally focus on some specific features of traffic as inputs, and they usually have a low potential to detect content-based attacks. However, it is well known that the content-based attacks, like SQL injection, malicious software, and viruses are the most destructive attacks against assets that are accessible on the Internet.

According to our study, only a few of previous \linebreak learning-based works on IDSs have considered content-based attacks. These works, like \cite{Wang2004PYLE}, \cite{Wang2006Anagram}, \cite{Perdisci2009McPAD}, and \cite{Cretu2008anomaly}, use $n$-gram methods for extracting the frequencies of characters in deterministic windows. However, as shown in \cite{Song2010ngramattack}, $n$-gram methods are vulnerable to mimicry attacks. In these kinds of attacks, some unused parts of packets like IP options or PADDING parts in exploits can be used for adjusting the frequencies of $n$-grams. 

A severe obstacle for analyzing the contents of network traffic is the large dimension of payloads. Nowadays, this challenge can be handled effectively by employing Deep Learning techniques \cite{Bengia_RepLrnReview_PAMI13,lecun2015deep}. In this paper, a deep learning-based intrusion detection method, called deep intrusion detection (DID) system is proposed. It uses the pure content of traffic (\ie, packet payload) as the input data. In the pre-processing phase, the content of each flow is converted to a numerical matrix. The learning and detection phases use this matrix for separating normal traffic from the malicious one. 

In this work, our primary contribution is to use all content bytes of traffic during the learning and detection phases. This goal is achieved by employing deep learning methods (in particular, in this work, we leverage using the LSTM neural network). Besides, we propose an appropriate pre-processing phase for feeding the traffic flows into the learning models. There are many studies around using deep learning models in IDS scope, as are reviewed in this paper. Still the main novelty of this paper is the use of the enriched raw content bytes of flows (not pre-extracted features) as the input, and the ability to distinguish the content-based attacks. Finally, we evaluate our proposed scheme on the CIC-IDS2017 dataset \cite{Sharafaldin2018ISCX}. This dataset has an appropriate variety of full captured normal and attacks traffic; in particular, it contains some content-based attacks like Heartbleed.

The remainder of the paper is organized as follows. In Section~\ref{sec:related}, we summarize the most relevant related works. Section~\ref{sec:method} presents the details of the proposed DID system. This system also includes a pre-processing phase for preparing contents of traffic flows to be fed to a deep learning model (\ie, an LSTM neural network). In Section \ref{sec:experiment}, the conducted experiments and results obtained are discussed. Finally, Section~\ref{sec:conclusion} concludes the paper and explains the possible future directions.

\section{Related Works} \label{sec:related}
In the following, we will review some of the learning-based approaches used in intrusion detection systems.

\subsection{Traditional Machine Learning Approach}
In the literature, various learning-based techniques such as support vector machine (SVM), naive Bayes, decision tree, random forest, and neural networks have been proposed for intrusion detection systems. 

SVM is one of the most popular classification algorithms used so far. It has been used in research studies like \cite{Heba2010Principle}, \cite{chen2009using}, \cite{jia2017application} and \cite{wang2017effective}. In this algorithm, the classification process is performed by detecting a set of hyperplanes, as separators, in a high-dimensional space. The high time-complexity of the learning phase and the difficulty of finding a suitable kernel function are the most important challenges of this method. Learning time complexity has a superlinear relation with the number of input instances. Besides, there is a quadratic relation between the size of the kernel matrix and the number of instances.

Bayesian classifiers \cite{heckerman1998tutorial} use Bayes' rule for predicting the membership of input data to classes. They are built by using expert knowledge or efficient algorithms that perform inference. In Naive Bayesian classifiers, features are assumed to be conditionally independent. Though this assumption is not satisfied in practice, however, experiments have proved its good performance. Many papers have used this technique, \eg, see \cite{Kruegel2003Bayesian} and \cite{Jemili2007Bayesian}.

Authors in \cite{Kruegel2003Bayesian} have suggested generating multi-Bayesian network models in which each one separately generates an anomaly score for the input traffic. In \cite{Jemili2007Bayesian}, an IDS based on Bayesian network classifiers is proposed. In this research, association rules are used for the detection of normal/intrusion traffic. New traffic will get a low probability level for each of the normal or attack groups. So, these suspicious connections will also be labeled as an attack. In the second phase, these attacks are classified into four known or unknown attack categories by Bayesian rules.

One of the main data mining techniques used in intrusion detection systems is associated with decision trees. In \cite{Kruegel2003decisiontree}, the misuse detection engine of Snort \cite{snort2018} is replaced by decision trees. Firstly, the existing rules are provided to a clustering algorithm to reduce the comparison needed to determine rules that are triggered by specific input data. These clusters are based on the values of important features. When the clustering algorithm reaches a rule set for the given feature of the input data, the decision tree determines the triggered rules inside that cluster. 

Random forests (RF) \cite{Breiman2001RF} consist of a collection of decision trees. In addition to good performance in comparison with SVM and neural networks (NNs), this approach can run efficiently on large datasets with many features. RF is robust against overfitting and can handle unbalanced data. Works like \cite{Zhang2008RF} and \cite{farnaaz2016random} use this technique.

Artificial neural networks (ANNs) were the most popular models used until the 1990s when SVM was invented. One of the benefits of SVM against ANN is its lower learning time besides having a less local minimum problem. However, with the emergence of new ANN variants like recurrent and convolutional NNs, the ANNs have begun to be used again.

In \cite{Lippmann2000ANN}, a detector for finding attacks on Telnet is proposed. This system extracts 89 pre-defined keywords from the Telnet sessions. These keywords represent the suspicious actions or well-known attacks in Telnet. After extracting the distribution of these keywords, their statistics are given to a binary neural network. Finally, the instances recognized as attacks are given to a secondary NN, which determines the class name of the attack. They have finally obtained detection rates up to 80\%.  

ANNs can also be used for the detection of DoS attacks like SYNFLOOD, UDPSTORM, and SMURF (for example, see \cite{Palagiri2002ANN}). For this purpose, authors of \cite{Palagiri2002ANN} use a time window, which is then labeled as normal or attack traffic. Since the input size of an ANN is fixed, they use a pre-processing phase with the aid of an anomaly-based ANN, namely, a self-organization map (SOM). SOM can cluster the input data into a fixed number of clusters. Hence, independently from the number of packets in the time window, a fixed number of inputs is provided for the ANN by this clustering technique. The model is evaluated by DARPA 1999 dataset \cite{Lippmannm2000DARPA99} and reaches 100\% detection of normal traffic and 76\% false-positive rate for attacks.

In 2017, feature reduction techniques had been proposed by using ANNs \cite{manzoor2017feature}. The authors use a combination of information gain and correlation for feature selection. Then, after normalizing the numbers of each class in the KDD99 dataset \cite{KDD99},  their model achieves the average recall value of 91.72\%.

\subsection{Deep Learning Approach}
The recurrent neural network (RNN) is a class of ANNs in which nodes have some amount of memory. As a result, in addition to the current input, the previous inputs can also influence the current output. These networks are suitable for sequential inputs that possess a dependency with each other. Long short-term memory (LSTM) network is a class of RNNs \cite{hochreiter1997LSTM}. LSTM has been proposed to solve the vanishing and exploding gradient by introducing some gates to the neural network structure. Therefore, LSTM can effectively learn the relations between items that are far away from each other in a sequence. Computer network flows, consisting of packets, form a sequence of data; hence, RNN and LSTM are natural candidates for analyzing of computer network traffics.

Authors of \cite{Agarap2018GRU} have employed gated recurrent unit (GRU), which is a variant of LSTM. They have slightly modified GRU and used SVM as a classifier instead of the softmax function. The goal of this modification is to increase the computational efficiency of the model. They have evaluated the proposed model with 2013 network traffic data obtained by the honeypot systems at Kyoto University. The inputs of this model are 24 statistical features of the dataset. For improving performance and reducing the computation cost of the mode, the continuous features are converted to bins and, finally, are represented in a one-hot format. Their model has an average accuracy of 80.53\%.

In another work, Kim et al. \cite{Kim2016LSTM} have applied LSTM architecture in IDS and use the KDD99 dataset for evaluating their proposed model. Their input vector contains 41 normalized features, and the output vector is composed of 4 attack classes and one non-attack class. In their evaluation, the average values of recall and fall-out are 98.79\% and 10\%, respectively. 

In \cite{Tang2016SDN}, the authors use deep learning for detecting anomalies in a software-defined network (SDN) environment. They use six basic features of the NSL-KDD dataset (duration, protocol type, SRC bytes, DST bytes, count, and SRV count) to detect anomaly flows. Finally, the attack detection accuracy is reported as 75.75\%.

Also, in some other research studies like \cite{Aminanto2018SAE}, \cite{Javaid2016AE}, and \cite{Aminanto2017SAE}, the deep learning approach is employed for the reduction of input dimensions by selecting among pre-extracted features. 


The authors of \cite{khan2019scalable} propose a scalable hybrid IDS with two-stages: the first stage is the anomaly detector module implemented by Spark ML traditional machine learning models; the next stage is the misuse detector, which is based on the Conv-LSTM network. Their evaluation is based on ISCX IDS 2012 dataset with 10-fold cross-validation tests. The results show a $97.29\%$ detection rate of attacks and a $0.71\%$ of false alarm rate.


Although deep learning methods have been proposed for solving intrusion detection problems so far, to the best of our knowledge, they use extracted features of inputs, as in traditional approaches. These features mostly represent general aspects of traffic flow, like \linebreak source/destination port/IP address, duration time, start time, and packet/byte number of sent or received packets. These features are generally crucial to the detection of some kinds of attacks like DDoS and portscan. However, many important attacks, like SQL injections, worms, viruses, and XSS, which are content-based attacks, have general features very similar to benign traffics. In the following, some traditional research studies which have paid attention to these kinds of attacks are reviewed.

\subsection{Content-Based Approach}
Generally, some restricting extracted features are used in machine learning-based intrusion detectors. These general features are rarely based on contents transmitted through the established flow. Consequently, content-based attacks have a high impact on the security and privacy of network applications and services in such systems.

In the following, we review some related works on content inspection for intrusion detection. Most of the payload-based detectors extract statistical features by using the $n$-gram technique. PYLE \cite{Wang2004PYLE}, Anagram \cite{Wang2006Anagram}, and McPAD \cite{Perdisci2009McPAD} are among the most well-known works. PYLE uses $1$-gram method and extracts the frequency of values in each byte of the packet. Anagram uses $5$-gram and stores the extracted $5$-grams in Bloom filters. There are two kinds of Bloom filters in this work: one designed for attacks and the other for benign $n$-grams. Finally, these two Bloom filters examine the input traffic.

It is evident that in $n$-gram analysis, the dimension of feature space grows dramatically. Hence, limited by the curse of dimensionality problem, in practice, this approach can be used at most for $n=2$, which yields 65536 features. To mitigate this problem, McPAD \cite{Perdisci2009McPAD} measures the frequency of the occurrences of pairs of symbols (bytes), which are $k$ bytes apart from each other in the payload. In this way, some information in $n$-grams with $n > 2$ can be extracted by such pairs of bytes. Moreover, this method will only generate $256^2$ features regardless of the value of $k$. 

In \cite{Song2010ngramattack}, the authors show that blending attacks can defeat $n$-gram methods. These attacks fill unused parts of network traffics with new characters in proportion to the target frequency and, consequently, convert the statistics of characters to become similar to benign traffics. Their evaluation shows that to launch an attack against a $5$-gram detector, at least two packets (\ie, about 2000 bytes) are needed. Besides, they propose fragmentation overlapping for solving larger values of $n$. Different operating systems (OSs) have different behaviors for extracting bytes in overlapping situations. They may prefer the first or last arrived overlapped bytes. The other bytes will be ignored by the OS. So these ignored bytes can be used in higher values of $n$ for deluding the $n$-gram detectors.

In another research \cite{Soheily2017ISCX}, after encoding the content by Base64, the integer values are extracted. Finally, the frequencies of these integer values are enumerated. Even though authors do not mention an $n$-grams method, but in fact, they use a $1$-gram approach.


Another related work is \cite{basumallik2019packet}. This paper focuses on false data injection attacks (FDIA) on phasor measurement units (PMU), which are utilities for monitoring power systems. In this paper, CNN is compared with RNN, LSTM, and traditional classifiers such as SVM. PMU packet data consists of $d$ different instances of data items, including $n$ univariate voltage and current phasor data stream. Finally, this work proposes a CNN model with 2 CNN layers, a dropout probability of 0.5, and a fully connected layer with 512 neurons, which achieves a $98.67\%$ accuracy.

\section{METHODOLOGY}
\label{sec:method}
The high dimensionality of traffic content is one of the biggest challenges in the detection of content-based attacks. Although this challenge can be addressed by employing deep learning methodology, according to our survey, all the previous proposed studies have focused on pre-extracted features which are vulnerable to content-based attacks.

\begin{figure}[h!]
	\centering
	\includegraphics[width=\linewidth]{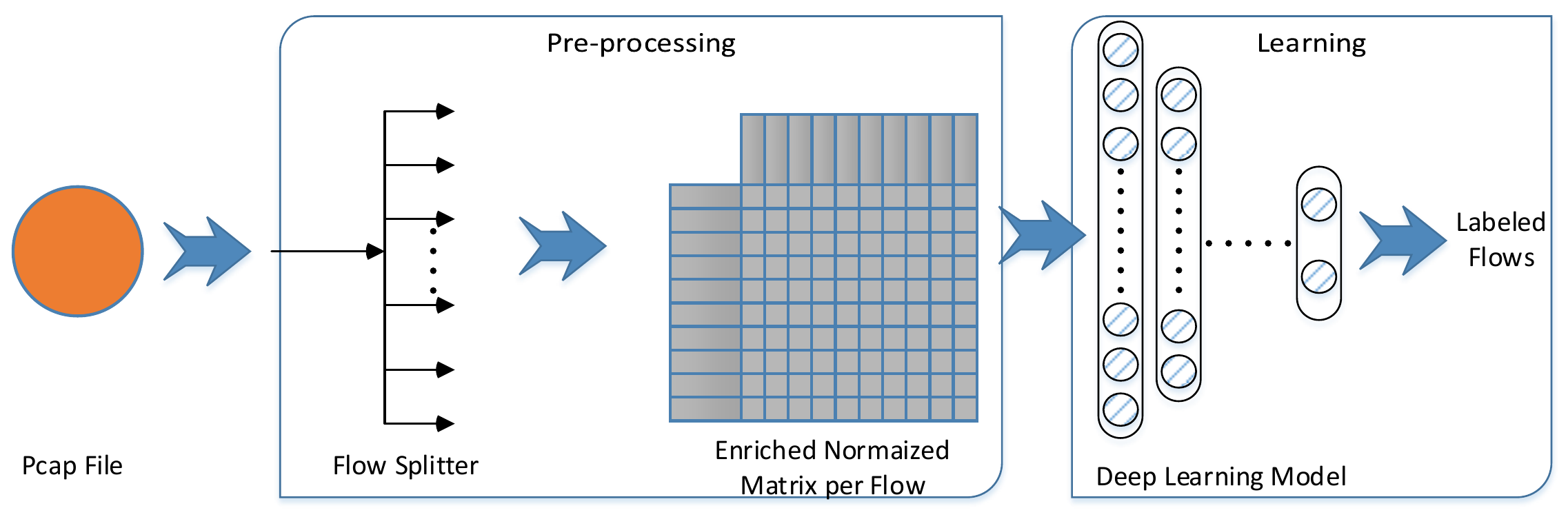}
	\caption{General illustration of DID system in the passive mode.}
	\label{fig:architecture}
\end{figure}

In this work, we propose a deep learning-based IDS method to extend the detection scope by covering the content-based attacks as well. Since traffic contents can have long-time dependencies, input feature space should have a high dimension. As deep learning methods are designed for such large data spaces, we propose using deep learning techniques directly on the raw bytes of contents instead of applying it to the extracted traffic features. The proposed method is called deep intrusion detection (DID). This method can be applied to both passive and on-line traffic. In this research, the passive mode is followed, as illustrated in Figure \ref{fig:architecture}.

Since traffic flows consist of sequences of data, algorithms like RNN and LSTM that are developed for sequential data are among the best candidates for the DID approach. In the following, we describe the proposed DID approach and explain how it uses deep learning methods to detect content-based attacks. In the following, we have two main subsections. The first one provides a complete description of the pre-processing module, and the second highlights the deep learning module of DID. In particular, in this work, we will employ LSTM for the deep learning module of DID. However, it should be noted that DID is not limited to use LSTM, rather, other algorithms can also be employed in the deep learning module.

\subsection{Pre-processing Phase in Deep Intrusion Detection}
\label{sec:method:DID}
Traditional learning methods highly depend on the pre-extracted features. As a result, the accuracy of such algorithms depends heavily on the selection of input features. Hence, these features should be found and extracted by experts, which makes the process expensive, time-consuming, and prone to error. Moreover, due to the increase of variant of known attacks and the emergence of new ones, extracting some static and definite features cannot provide adequate information for intrusion detection tasks.

In contrast, deep learning algorithms can extract complicated features from the raw data automatically. Consequently, to address the above issues, DID uses deep learning techniques to learn various cyber-attacks, including content-based attacks. It is well known that deep learning algorithms can detect sophisticated relations in high dimensional spaces. Hence, they are good candidates for the detection of content-based attacks. 

Although some content-based intrusion detection \linebreak systems like \cite{Wang2004PYLE}, \cite{Wang2006Anagram}, and \cite{Perdisci2009McPAD} focus on packet-level granularity, in real-world, some packets can belong to both benign and attack flows (\eg, SYN or FIN packets, or HTTP GET requests in DDoS attacks). Moreover, some attacks are distributed among more than one packet. Therefore, the concept of malicious traffic resides in the flow contents. As a result, we assume that the input to the DID method is based on flows instead of packets.

\subsubsection{Basic Normalized Matrix}
In this work, we propose an offline version of DID, where each flow is considered as an input sequence to an LSTM neural network. Each packet represents a data point in the input sequence. Since the maximum Ethernet frame size is around $1514$ bytes, so we consider $1514$ as the dimension for each packet. Hence, the input is assumed to be a sequence of $1514$-dimensional points.

Additionally, the size of input sequences depends on the number of packets in the traffic flows. In the offline DID, we assume some reasonable maximum value for the number of packets (which we will later determine this parameter by inspecting the dataset). Finally, since each byte is in the range of $0$ to $255$, in order to improve the deep network performance and make the parameters on the same scale, we normalize each byte value to a number between $0$ and $1$, by dividing it to $255$.

\begin{figure}[h!]
	\centering
	\includegraphics[width=0.8\linewidth]{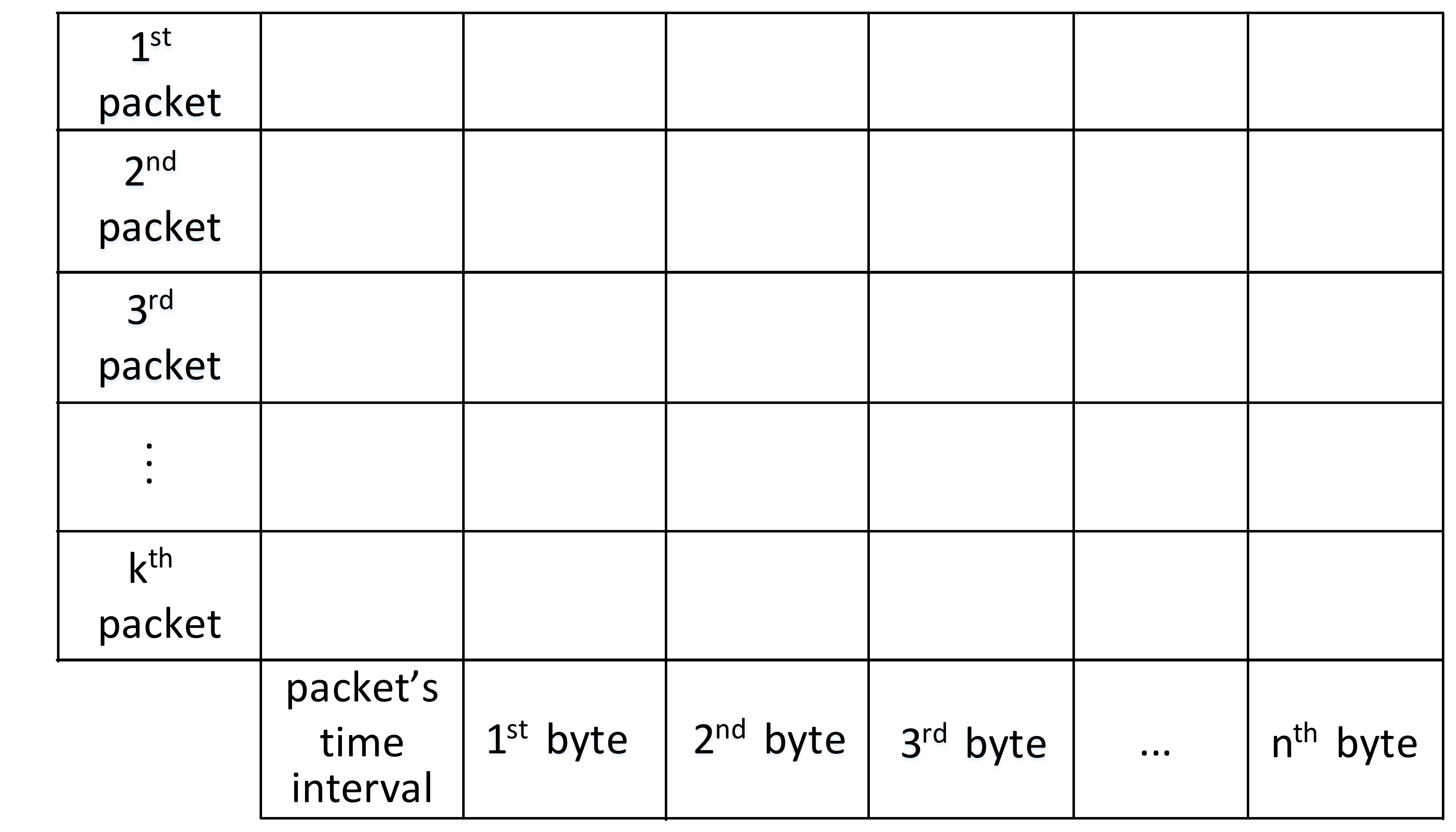}
	\caption{The output of the pre-processing phase in the form of a normalized matrix fed into the deep learning module. Each cell, $C_{ij}$, represents the content of the $j^{th}$ byte of the $i^{th}$ packet. The first column is different from other columns and represents the time interval between $i^{th}$ packet and $(i-1)^{th}$ packet.}
	\label{fig:basic-normalized-matrix}
\end{figure}

According to the pre-processing phase explained above, we have a normalized matrix per each flow (as depicted in Figure \ref{fig:basic-normalized-matrix}), where rows describe different packets in the flow, and the $i$th column contains the normalized value of the $i$th bytes of packets. Moreover, we add a column to the matrix for storing inter-arrival times of packet flows to detect attacks such as HTTP flooding, which sends some benign requests continuously over the established connection. These normalized matrices can be the input of the deep learning module of DID (we will later enrich these matrices).

Considering a large number of parameters in deep learning algorithms, and a limited number of flows that are used in the training phase, there is a reasonable chance of overfitting if the datasets are not used with enough care. As an example, IP addresses can be a misleading factor. This misleading effect exists in most available public datasets like CIC-IDS2017 \cite{Sharafaldin2018ISCX} and KDD99 \cite{KDD99}. In \cite{Mahoney2003DARPA}, authors have shown that many parameters of the DARPA 99 \cite{Lippmannm2000DARPA99} traffic, like TTL (Time To Live), ToS (Type of Service), and the IP addresses, can cause overfitting. For example, TTLs of the attack traffics are mostly 126 and 253, but benign traffic has nine restricted values, which are different from the attack ones. Besides, source IP addresses of attacks are different from benign traffics and can simply be used for discrimination. The KDD99 dataset also has inherited these vulnerabilities. Since the attack traffic constitutes a small part of the dataset, there are so many IP addresses that are purely normal, and the algorithm can assign a substantial weight for IP addresses to attain higher accuracy. However, we know that this is not a valid assumption in the real-world. 

Considering the above issues, in our pre-processing phase, we eliminate some bytes of packets that belong to fields like CHECKSUM and IP addresses. Specifically, the total data-link header, the Source/Destination IP addresses and the checksum from the network header, and the checksum of the transport layer are the removed items in the pre-processing phase. It should be noted that this elimination can cause some performance reduction in the detection phase. For example, ignoring the client's IP address in a monolithic environment, like a university, can avoid overfitting, but in heterogeneous networks with different types of clients, some valuable information can be missed. Besides, server IP addresses can be beneficial in server-side IDSes. So, in the real world, this elimination should be applied according to the conditions of the deployment environment.

\subsubsection{Enriched Normalized Matrix}
The pre-processing phase can be completed by enriching the normalized matrices. The basic pre-\linebreak processing matrices are adequate for detecting flow-based attacks. However, other kinds of attacks can be recognized by considering some intra-flows features. These features are also added to the first row of the basic pre-processing matrix to make it richer. For example, \emph{flooding} attacks can be generated by making many legitimate connections rapidly, and these kinds of attacks can be detected by adding time intervals between flows. Since in the real world, the normal and attack flows are interleaved, the computation of the time interval between flows should be based on the original flows' arrival times. The other approach is based on splitting the flows into benign traffic and attack, and then extracting the time interval in each subgroup. This approach can increase the detection error ratio when there exists normal traffic between attacks.

To address the intra-flow attacks, we use four more extensive intra-flows features as follows: aggregative \linebreak source or destination address repetition in a fixed-size bucket of packets or in a time window. Attacks like DDoS use multiple different IP addresses to send requests to the victim server, called Type I attacks. Detection of this kind of attack can be done by aggregating flows that have the same destination IP address. On the other hand, in some attacks like port scanning, a single client IP address tries to recognize different active services (ports) on a specific victim IP address or a specific service (port), which is activated on a network range. The mentioned scenarios have the same source address and destination port or same source and destination address, respectively. For simplification, we call these kinds of attacks as Type II.

Another important aspect of detecting intra-flow attacks is network bandwidth. For networks with low bandwidth, a fixed size window (or bucket) is used for aggregation. As the time interval between flows can exceed the time threshold, time windows cannot detect the attacks. On the other hand, the fixed-size window cannot detect attacks in high-speed traffics because the window will be filled rapidly, and the new information will overwrite the older ones. A time window can handle this situation as well. In real networks, bandwidth has no fixed value, and according to the conditions like days vs. nights, it can have low or high bandwidth. So we use a combination of these two kinds of windows for the detection of intra-flow attacks. 

The four aforementioned intra-flow features are extracted per each flow. Detection of Type I attacks depends on the aggregation of flows based on their destination IP addresses. Hence, as a new flow arrives, it is compared with flows that are observed in the fixed-size and fixed-time windows. The number of flows having the same IP address as the new one in both windows is used as features. Similarly, aggregation based on the source IP address is done for the detection of Type II attacks. In this case, the source address of each flow is compared with the source addresses of flows in the fixed-size and fixed-time windows. 

\begin{figure}[h!]
	\centering
	\includegraphics[width=1.0\linewidth]{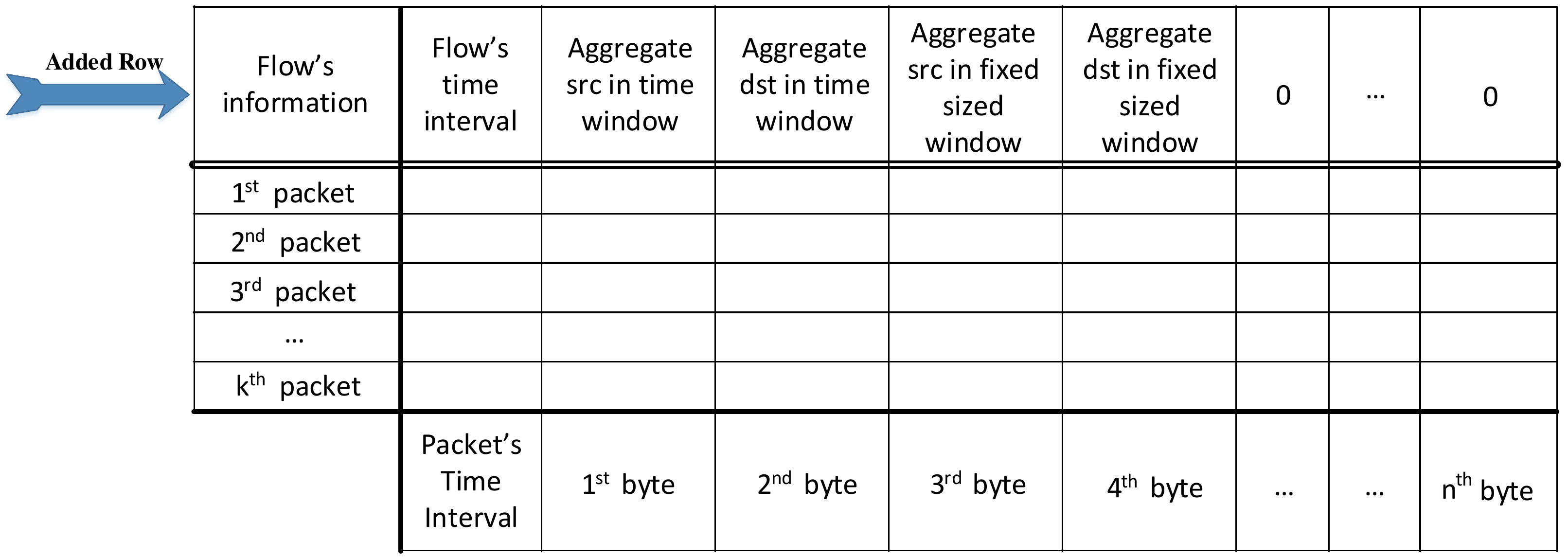}
	\caption{The structure of enriched normalized matrices used as the input to the deep learning module of DID. The main difference with Figure \ref{fig:basic-normalized-matrix} is the first row of the matrix, added as the features that represent the intra-flow context.}
	\label{fig:reached2-normalized-matrix}
\end{figure}

Finally, the five new intra-flow features will be added to the first row of the basic normalized matrix (see Figure \ref{fig:reached2-normalized-matrix}). This enriched matrix will be used as an input of the deep learning module of DID. In the following, some candidates for DID deep learning module are discussed, and the LSTM model is implemented.

\subsection{Deep Learning Module of Deep Intrusion Detection}
As mentioned earlier, in DID, we prepare a rich normalized matrix as the input for a deep learning algorithm. This matrix has the potential for extracting content-based and some intra-flow attacks. In the following, some candidates for deep learning modules are discussed. The main important point, which is common among the proposed methods, is the sequential nature of these algorithms. In fact, since packets, flows, and network traffics are all, in general, sequential data, the chosen algorithms should match or benefit from this feature.  

\subsubsection{Recurrent Neural Networks}
Recurrent Neural Network (RNN) is suitable for learning patterns in data sequences and time series, such as processing natural languages and genetic data \cite{dorffner1996neural}. This feature makes RNN an extremely useful tool for analyzing computer network traffic. The difference between recurrent neural networks and feed-forward neural networks is that besides the current input, some information from previous inputs is also processed. In RNN, decision making related to an input instant at the moment $t$ depends on the decision made at the moment $t - 1$.

The mathematical definition of the forward memory transfer process in recursive neural networks is as follows
\[
h_t = \phi (Wx_t + Uh_ {t-1}),
\]
where $ h_t $ is the state of the hidden layer of the recurring neural network at the moment $t$. The value of $ h_t $ is a function of the input at the moment $t$ (\ie, $ x_t $) which is multiplied to hidden layer weights $ W,$ and the last moment hidden layer feedback $h_{t-1}$ which is multiplied to its own weights $U$. The weight matrices apply the relative importance of the input at the current moment and the feedback input from the previous moment.

\subsubsection{LSTM}
\begin{figure}
	\centering
	\includegraphics[width=\linewidth]{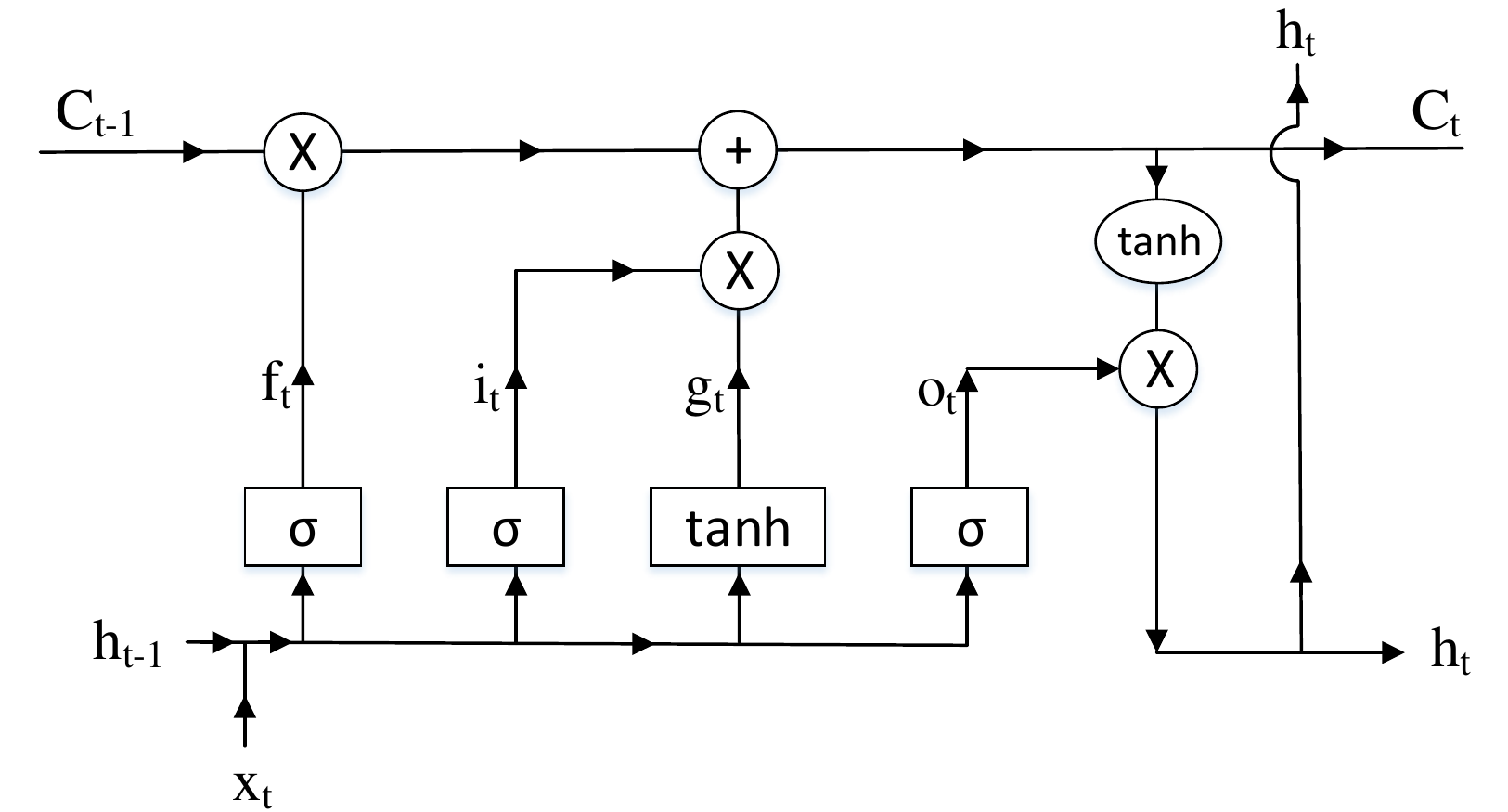}
	\caption{The internal structure of an LSTM cell.}
	\label{fig:lstm}
\end{figure}

LSTM is a special type of recurring neural network which is capable of learning long-term dependencies. These networks have proven to be very effective in many different circumstances and are now widely used in practice. An LSTM layer consists of some similar units, called LSTM cell.  Inside each cell, four neural networks are linked to each other in a specific structure (see Figure \ref{fig:lstm}). This special structure enables an LSTM network to learn simultaneously short and long-term dependencies very well. For more details on the LSTM the interested readers can refer to \cite{Goodfellow-et-al-2016}.



\subsubsection{LSTM-Based Classifier}
Since, in practice, it has been observed that LSTM based classifiers and their variants perform very well on sequential data, we construct a deep learning model with two LSTM layers as a proof of concept for our proposed deep intrusion detection (DID) framework. The hyper-parameters that are dedicated to this model are based on the evaluations which are discussed later. 

\begin{figure}[h!]
	\centering
	\includegraphics[width=\linewidth, keepaspectratio]{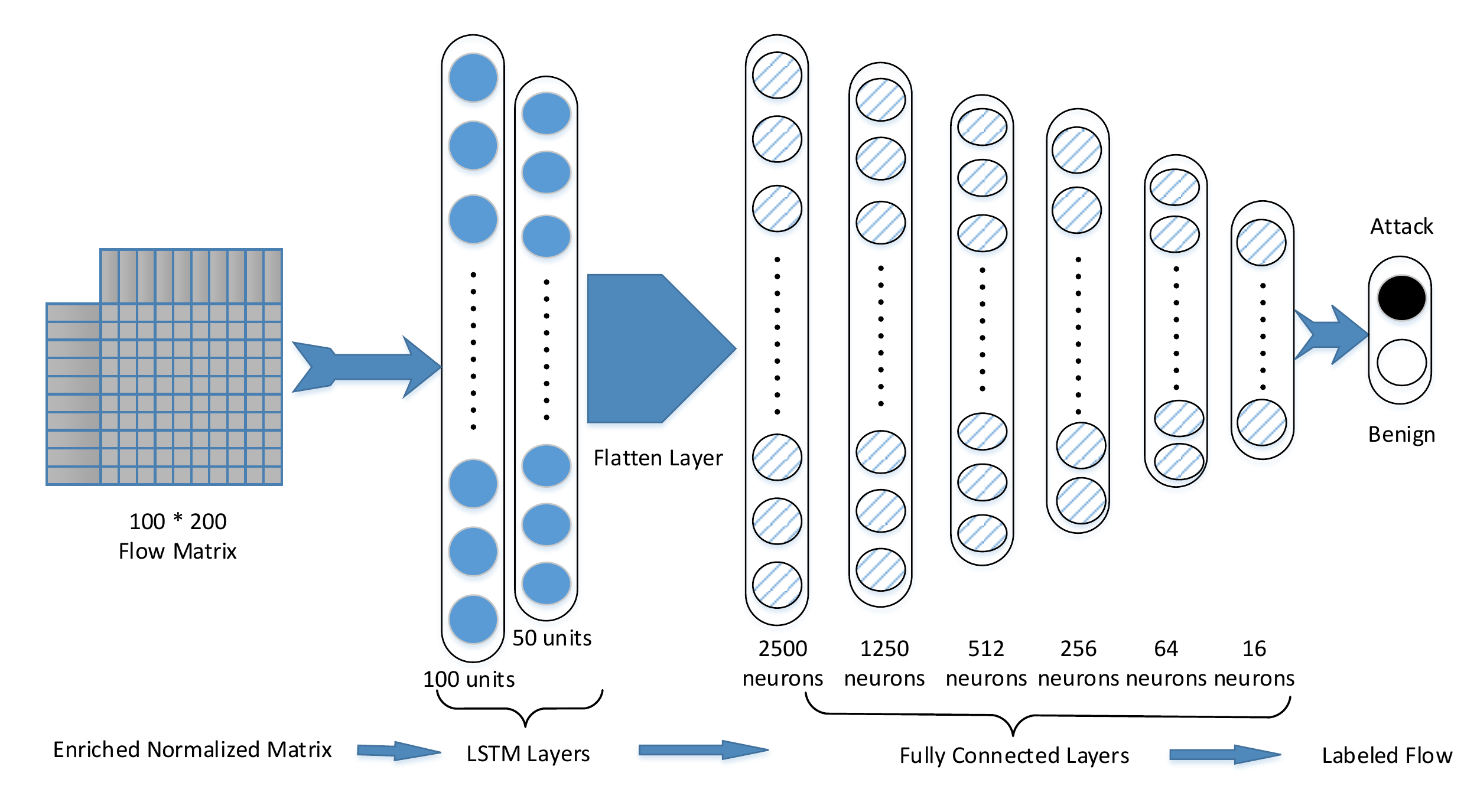}
	\caption{The proposed LSTM-based classifier used in the our DID framework.}
	\label{fig:lstm-classifier-details}
\end{figure}

Figure~\ref{fig:lstm-classifier-details} presents the details of the proposed model. As shown in Figure~\ref{fig:lstm-classifier-details}, after extraction of sequential features with LSTM layers (with 100 and 50 units, respectively), some fully connected layers (with 2500, 1250, 512, 256, 64, and 16 neurons) are employed to extract the more complicated features. Finally, a softmax layer is applied for binary classification between attack and benign traffics. The activation functions of all layers (except the last one) are ReLU, and in order to avoid overfitting, some dropout layers with a 20\% drop rate are added among fully connected ones. Finally, the Adam algorithm is used for optimization in the training phase, and the loss value is computed by binary cross-entropy as the loss function.

\section{EXPERIMENT}
\label{sec:experiment}
This section contains a real implementation of a DID instance on the CIC-IDS2017 and CSE-CIC-IDS2018 datasets. In the following, first, the CIC-IDS2017 and CSE-CIC-IDS2018 datasets are briefly introduced. Then, after explaining the pre-processing phase, the experimental results are presented and compared with previous works. The reported results in this paper are based on the 10-fold cross-validation tests.

\subsection{Dataset}
\label{sec:method:dataset}
In this work, we use the CIC-IDS2017 and CSE-CIC-IDS2018 datasets to benchmark the proposed DID method. According to \cite{ferrag2020deep}, there are few labeled datasets with PCAP format traffic files; among them, ISCX/CIC IDS is one of the best and up-to-date ones. CIC IDS is the only option with adequate labeled attacks/benign traffics in PCAP format. Hence, we made our evaluations based on this dataset solely. This dataset is made of 50 GB network traffic captured in five different days, which is the most recent IDS evaluation dataset and contains different types of attacks. Especially, content-based attacks like Heartbleed are also included in this dataset. Traffic capturing is done in a simulated computer network with several servers and clients. The developers of CIC-IDS2017 have analyzed real traces of a client-server network and have tried to create the same profile for the clients. The details of the five days of network traffic are shown in Table~\ref{table:1}. 

\begin{table}
	\caption{Details of the CIC-IDS2017 dataset.}	\label{table:1}
	\centering
	\resizebox{\linewidth}{!}{%
		\begin{tabular}{|c|c|c|c|}
			\hline
			\parbox[c][0.4cm][c]{2cm}{\centering Day} & Attack type & Attack Size & Benign Size\\
			\hline\hline
			\parbox[c][0.4cm][c]{2cm}{\centering Monday} & - & 0B & 11GB\\
			\hline
			\parbox[c][0.4cm][c]{2cm}{\centering Tuesday} & Brute Force & 51MB & 11GB\\
			\hline
			\parbox[c][0.4cm][c]{2cm}{\centering Wednesday} & DoS / DDoS & 2GB & 11GB\\ 
			\hline
			\parbox[c][0.4cm][c]{2cm}{\centering Thursday}& Web Attack, Infiltration  & 42MB & 8.4GB\\
			\hline
			\parbox[c][0.4cm][c]{2cm}{\centering Friday}& Botnet ARES, Port Scan  & 2GB & 7.5 GB\\
			\hline
		\end{tabular}
	}
\end{table}

The main advantages of this dataset compared to the previous ones are:
\begin{itemize}
	\item Implementing a complete network configuration, including Modem, Firewall, Switches, Routers, and a variety of operating systems.
	\item Simulation of user profiles.
	\item The dataset is labeled. This is a requirement for classification purposes. Besides, it presents the full captured traffic without anonymization techniques.
	\item Implementing all kinds of interactions in the network.
	\item Using a wide range of protocols and network attacks.
\end{itemize}

Although, as explained above, this dataset has many advantages, it has its shortcomings too. One of the most important deficiencies of this dataset is its limited variety of protocols and attacks compared to real-world traffics. For example, IP addresses of attack traffic are very limited, and hence, the other IP addresses can be recognized as pure benign traffic. More precisely, attacks on Tuesday and Wednesday are just focused on one and two destination addresses, respectively.  Besides, DoS attacks on Friday are all from a specific client IP address. On the other hand, real network conditions like packet loss and different TTLs are not presented in this dataset. Moreover, so many kinds of applications like social networking are not considered.

For reporting more reliable results we also evaluated the DID over a newer version of the IDS evaluation dataset which is called CSE-CIC-IDS2018 \cite{CIC2018} and is introduced by CIC (Canadian Institute for Cybersecurity) and CSE (Communications Security Establishment) collaboratively. This dataset extends the variety of packets, OSes, network topology, and servers. For example, the attacking infrastructure includes 50 machines and the victim organization has 5 departments includes 420 PCs and 30 servers. However, the attack types and normal protocols remain the same as the CIC-IDS2017 dataset. Due to its better implementation topology, this dataset can provide a better challenge for the DID framework.

Finally, as mentioned above, according to our survey, the CIC-IDS2017 and CSE-CIC-IDS2018 are the best datasets available in the context of IDS. However, we should be aware of their weaknesses and simplicity. Obviously, in the real world, we need to implement more sophisticated ML models with a higher number of layers and nodes in each layer. Moreover, as discussed in Section~\ref{sec:method:DID}, we should be aware of the lack of diversity of these datasets in the pre-processing phase.

\subsection{Pre-processing}
In this phase, we prepare the dataset as the input of a neural network. First, we need to extract and split the network flows from the pcap files. To this end, we read the large pcap files and make separate files per each flow. Flow separation is based on the source port, destination port, source IP address, destination IP address, and flow start time. The end of flow is reached when the TCP FIN packet is read, or the maximum flow time (1,200,000 ms) is passed.

\subsubsection{Constructing the Input Matrix}
Network flows are not suitable to be input directly to the neural network. To make the flows applicable, we have to apply several changes to them. First, we read the frames of each flow. The data link header is removed for extracting the packet since it does not have any information for network intrusion detection tasks. Then we read the bytes of the packet and divide them by 255 to obtain a normalized value between 0 and 1.

The maximum size of each packet is 1514 bytes, and smaller packets are padded by zero-value bytes. Besides, since the header length of UDP is less than TCP, we add zero to the end of the UDP header so it will have the same size as TCP header. There are some fields in network traffic, which can mislead the deep learning model. For example, the checksum field can have random values, and most probably, it is useless. Moreover, as explained above, IP addresses can lead to the overfitting problem.  We mask the value of these fields by zero. In the end, we will have an $n\times1514$ matrix, where $n$ is the number of packets.

\begin{figure}[h!]
	\centering
	\includegraphics[width=\linewidth, keepaspectratio]{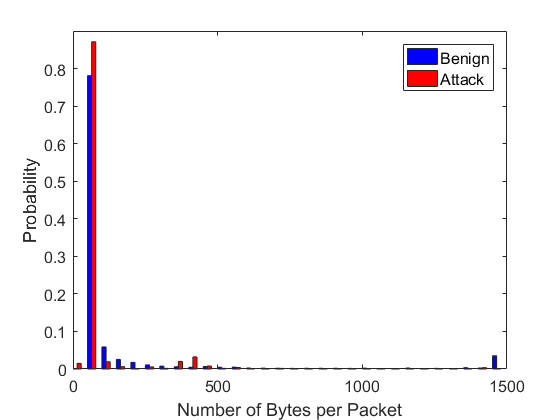}
	\caption{The distribution of number of bytes per packet in the benign and attack traffics.}
	\label{fig:bytes-per-packet}
\end{figure}

\begin{figure}[h!]
	\centering
	\includegraphics[width=\linewidth, keepaspectratio]{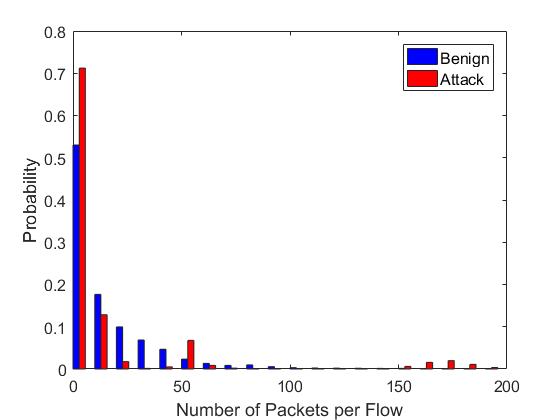}
	
	\caption{The distribution of the number of packets per flow in benign and attack traffics.}
	\label{fig:packets-per-flow}
\end{figure}

The dimensions of the input matrix for this dataset can be reduced by inspecting the dataset traffic. As shown in Figure~\ref{fig:bytes-per-packet}, packet size in normal and attack traffic has two distinct ranges: packets with only the first 200 bytes, and packets with the maximum size of 1514 bytes. By performing several experiments on the dataset, we found that the first 200 bytes of each packet constitutes the discriminant bytes, and inspecting extra bytes has no significant impact on the learning accuracy. In addition, benign and attack flows in this dataset contain mainly less than 100 packets (as shown in Figure \ref{fig:packets-per-flow}). So, inspecting only the first 100 packets of each flow can yield almost a complete evaluation of the nature of flows in the ISCX/CIC 2017 dataset.

Finally, we have chosen the first 200 bytes of the first 100 packets of each flow as an input matrix according to the nature of flows in this dataset.

\subsubsection{Subsampling}
As shown in Table \ref{table:1}, the size of the CIC-IDS2017 dataset is 50GB before the pre-processing phase, and the pre-processing phase increases its size tremendously to more than 500GB. Due to hardware limitations, we cannot use all traffic flows to train the neural network. Therefore, we need to reduce the size of the dataset. Also, the dataset is imbalanced, and the number of benign flows is much higher than the number of attack flows. This imbalance in data does not allow to train the neural network correctly. To fix these issues, we choose all of the attack flows and randomly select the same amount of benign flows, balancing the dataset and reducing the input data size. Finally, we have a pre-processed dataset with a size of around 40GB. 

The size of the CSE-CIC-IDS2018 dataset is even more challenging. Its original size is around 480GB and captured in 10 different days which is represented in Table \ref{table:ids2018}. Table \ref{table:binary-balance} represents the selected flow numbers of each category in these two datasets. The categories and their sub-category attacks are as follows: Botnet, Port Scan, DoS/DDoS (DoS slowloris, DoS Slowhttptest, DoS Hulk, DoS GoldenEye), Heartbleed/Infiltration, Brute Force (FTP-Patator, SSH-Patator), Web Attack (Brute Force, XSS, SQL Injection). The two datasets have the same attack categories with some little differences such as The 2017 version contains heartbleed attacks. On the other side, the 2018 version has lots of infiltration attacks. Besides, the port scan attacks are absent in the CSE-CIC-IDS2018 and the number of its web attacks is more limited.

\begin{table}
	\caption{Details of the CSE-CIC-IDS2018 dataset.}	\label{table:ids2018}
	\centering
	\resizebox{\linewidth}{!}{%
		\begin{tabular}{|c|c|c|}
			\hline
			\parbox[c][0.4cm][c]{4cm}{\centering Day} & Attack type & Pcap size\\
			\hline\hline
			\parbox[c][0.4cm][c]{4cm}{\centering Friday-02-03-2018} & Botnet  & 45GB\\
			\hline
			\parbox[c][0.4cm][c]{4cm}{\centering Friday-16-02-2018/} & DoS & 39GB\\
			\hline
			\parbox[c][0.4cm][c]{4cm}{\centering Friday-23-02-2018} & Web Attacks & 59GB\\ 
			\hline
			\parbox[c][0.4cm][c]{4cm}{\centering Thursday-01-03-2018}& Infiltration & 53GB\\
			\hline
			\parbox[c][0.4cm][c]{4cm}{\centering Thursday-15-02-2018}& DoS & 41GB\\
			\hline
			\parbox[c][0.4cm][c]{4cm}{\centering Thursday-22-02-2018}& Web Attacks   & 50GB\\
			\hline
			\parbox[c][0.4cm][c]{4cm}{\centering Tuesday-20-02-2018}& DDoS  & 46GB\\
			\hline
			\parbox[c][0.4cm][c]{4cm}{\centering Wednesday-14-02-2018}& Brute Force  & 40GB\\
			\hline
			\parbox[c][0.4cm][c]{4cm}{\centering Wednesday-21-02-2018}& DDoS  & 55GB\\
			\hline
			\parbox[c][0.4cm][c]{4cm}{\centering Wednesday-28-02-2018}& Infiltration   & 53GB\\
			\hline
		\end{tabular}
	}
\end{table}

\begin{table}[h]
	\centering
	\caption{The balanced number of evaluated flows in the binary classification experiment.}
	\resizebox{\linewidth}{!}{%
		\begin{tabular}{|c|c|c|c|}
			\hline
			\parbox[c][0.5cm][c]{1cm}{\centering Category}                & \parbox[c][0.5cm][c]{2cm}{\centering Sub-Category}            & \parbox[c][0.5cm][c]{2cm}{\centering CIC-IDS2017} & \parbox[c][0.75cm][c]{2cm}{\centering CSE-CIC-IDS2018} \\ \hline
			\parbox[c][0.5cm][c]{2cm}{\centering Benign}                  & -                       & \parbox[c][0.5cm][c]{2cm}{\centering 15000}       & \parbox[c][0.25cm][c]{2cm}{\centering 15000}           \\ \hline
			\multirow{8}{*}{Attack} & \parbox[c][0.5cm][c]{2cm}{\centering Web Attacks}             & \parbox[c][0.25cm][c]{2cm}{\centering 1500}        & \parbox[c][0.25cm][c]{2cm}{\centering 200}             \\ \cline{2-4} 
			& \parbox[c][0.5cm][c]{2cm}{\centering Botnet}                  & \parbox[c][0.25cm][c]{2cm}{\centering 2000}        & \parbox[c][0.25cm][c]{2cm}{\centering 3000}            \\ \cline{2-4} 
			& \parbox[c][0.5cm][c]{2cm}{\centering Port Scan}               & \parbox[c][0.25cm][c]{2cm}{\centering 2000}        & -               \\ \cline{2-4} 
			& \parbox[c][0.5cm][c]{2cm}{\centering DoS / DDoS}              & \parbox[c][0.25cm][c]{2cm}{\centering 6000}        & \parbox[c][0.25cm][c]{2cm}{\centering 6000}            \\ \cline{2-4} 
			& \parbox[c][0.5cm][c]{2cm}{\centering Brute Force}             & \parbox[c][0.25cm][c]{2cm}{\centering 3500}        & \parbox[c][0.25cm][c]{2cm}{\centering 3000}            \\ \cline{2-4} 
			& \parbox[c][0.75cm][c]{2cm}{\centering Heartbleed / Infiltration} & \parbox[c][0.25cm][c]{2cm}{\centering 10}          & \parbox[c][0.25cm][c]{2cm}{\centering 3000}            \\ \hline
		\end{tabular}%
	}
	\label{table:binary-balance}
\end{table}

As mentioned above, the subsampling technique is one of the main approaches to pre-process an unbalanced dataset before input to ML models. As mentioned above, in the binary classification scenario, we have balanced the two classes (benign and attack). When we consider a multi-class classification setup, the same technique can be applied. In this case, the subsampling equalizes all classes, including the benign and other attack categories.

Finally, the two evaluated CIC datasets have the same attack types and categories. The attack types of each attack category and the selected number of flows for each attack category of the datasets are described in Table \ref{table:multi-balance}. 

\begin{table}[h]
	\centering
	\caption{The balanced number of evaluated flows in the multi-class classification experiment.}
	\resizebox{\linewidth}{!}{%
		\begin{tabular}{|c|c|c|c|}
			\hline
			\parbox[c][0.5cm][c]{1cm}{\centering Category}                & \parbox[c][0.5cm][c]{2cm}{\centering Sub-Category}            & \parbox[c][0.5cm][c]{2cm}{\centering CIC-IDS2017} & \parbox[c][0.75cm][c]{2cm}{\centering CSE-CIC-IDS2018} \\ \hline
			\parbox[c][0.5cm][c]{2cm}{\centering Benign}                  & -                       & \parbox[c][0.5cm][c]{2cm}{\centering 5000}       & \parbox[c][0.25cm][c]{2cm}{\centering 5000}           \\ \hline
			\multirow{8}{*}{Attack} & \parbox[c][0.5cm][c]{2cm}{\centering Web Attacks}             & \parbox[c][0.25cm][c]{2cm}{\centering 1500}        & \parbox[c][0.25cm][c]{2cm}{\centering 200}             \\ \cline{2-4} 
			& \parbox[c][0.5cm][c]{2cm}{\centering Botnet}                  & \parbox[c][0.25cm][c]{2cm}{\centering 2000}        & \parbox[c][0.25cm][c]{2cm}{\centering 3000}            \\ \cline{2-4} 
			& \parbox[c][0.5cm][c]{2cm}{\centering Port Scan}               & \parbox[c][0.25cm][c]{2cm}{\centering 2000}        & -               \\ \cline{2-4} 
			& \parbox[c][0.5cm][c]{2cm}{\centering DoS / DDoS}              & \parbox[c][0.25cm][c]{2cm}{\centering 6000}        & \parbox[c][0.25cm][c]{2cm}{\centering 6000}            \\ \cline{2-4} 
			& \parbox[c][0.5cm][c]{2cm}{\centering Brute Force}             & \parbox[c][0.25cm][c]{2cm}{\centering 3500}        & \parbox[c][0.25cm][c]{2cm}{\centering 3000}            \\ \cline{2-4} 
			& \parbox[c][0.75cm][c]{2cm}{\centering Heartbleed / Infiltration} & \parbox[c][0.25cm][c]{2cm}{\centering 10}          & \parbox[c][0.25cm][c]{2cm}{\centering 3000}            \\ \hline
		\end{tabular}%
	}
	\label{table:multi-balance}
\end{table}

\subsection{Experimental Results}
\label{sec:exresult}
After converting each flow to an enriched input matrix, we have split the dataset randomly into three subsets. The first set, which contains 64\% of the flows, is used to train and tune the weights of the deep learning model. The second and third sets are used during validation and test phases and contain 16\% and 20\% of flows, respectively. We performed 10-fold cross-validation and grid search for the hyper-parameter tuning.

There exist several metrics for evaluating the performance of the trained model. Among them, we have chosen precision (PR), recall (RC), fall-out (FO), and $F_{1}$ score ($F_{1}$). Based on a confusion matrix, equations of these parameters are stated as follows (TP: true positive, FP: false positive, TN: true negative, and FN: false negative)

\begin{align} 
	\mathrm{PR} &= \mathrm{TP}/(\mathrm{TP}+\mathrm{FP}), \label{eq:precision} \\
	\mathrm{RC} &= \mathrm{TP}/(\mathrm{TP}+\mathrm{FN}), \label{eq:recall}\\
	\mathrm{FO} &= \mathrm{FP}/(\mathrm{FP}+\mathrm{TN}), \label{eq:fall-out}\\
	F_{1} &= \frac{2\times \mathrm{PR} \times \mathrm{RC}}{\mathrm{PR}+\mathrm{RC}}. \label{eq:F1-score}
\end{align}

Recall (RC) is a valuable metric in IDSs as it determines the ratio of attacks that have been detected to the actual attacks. Besides, the ratio of benign flows, labeled as attacks, to the total actual benign flows is determined by the fall-out (FO). Precision (PR) shows the ratio of correctly generated alerts (existence of attacks) to all alerts. This metric represents the trust of network administrators to the generated security alarms. Finally, $F_{1}$ score tries to make a balance between the importance of precision and recall. This is achieved by calculating the harmonic mean of these valuable metrics.

To implement our deep learning model, we have used the Keras library \cite{chollet2017}, with Tensorflow \cite{tensorflow2015-whitepaper} as its backend. The characteristic of our experiment environment is shown in Table~\ref{tab:Hardware}.

\begin{table}
	\caption{The hardware specification of the experiment environment.} \label{tab:Hardware}
	\centering
	\resizebox{\linewidth}{!}{%
		\begin{tabular}{|c|l|}
			\hline
			\parbox[c][][c]{2cm}{\centering OS} & \parbox[c][0.9cm][c]{5cm}{Debian version 9.3 with kernel 4.9.0-amd64}\\
			\hline
			CPU & \parbox[c][0.9cm][c]{5cm}{Intel(R) Xeon(R) X5680 3.33GHz with 24 virtual cores}\\
			\hline
			RAM & \parbox[c][0.5cm][c]{5cm}{18 GB}\\
			\hline
			GPU & \parbox[c][0.5cm][c]{5cm}{GeForce GTX 1080 Ti}\\
			\hline
			GPU Frame Buffer & \parbox[c][0.5cm][c]{5cm}{11 GB}\\
			\hline
		\end{tabular}
	}
\end{table}

\subsubsection{Evaluation of the Traditional ML Models}
First, we evaluate the performance of different traditional machine learning classifiers like Support Vector Machine (SVM), Random Forests (RF), and Naive Bayes (NB) over both the CIC-IDS2017 and CSE-CIC-IDS2018 datasets. We have only presented the best results obtained by the hyper-parameters chosen through an empirical random search (see Table \ref{table:Traditional evaluation metrics}). The results are based on pre-extracted features, which are presented in CSV files alongside the dataset.

\begin{table}[h]
	\centering
	\caption{The traditional ML models for the binary classification evaluated on the pre-extracted features given in CSV files of CIC-IDS2017 and CSE-CIC-IDS2018.}
	\resizebox{\linewidth}{!}{%
		\begin{tabular}{|c|c|c|c|c|}
			\hline
			Dataset                          & \parbox[c][0.75cm][c]{1cm}{\centering Model}    & \parbox[c][0.75cm][c]{2cm}{\centering Precision} & \parbox[c][0.75cm][c]{2cm}{\centering Recall} & \parbox[c][0.75cm][c]{2cm}{\centering $F_{1}$ score} \\ \hline
			\multirow{5}{*}{CIC-IDS2017}     & \parbox[c][0.75cm][c]{1cm}{\centering SVM} & 0.79                                                                                    & 0.76                                                                                 & 0.75                                                                                        \\ \cline{2-5} 
			& \parbox[c][0.75cm][c]{1cm}{\centering RF}  & 0.98                                                                                    & 0.98                                                                                & 0.98                                                                                        \\ \cline{2-5} 
			& \parbox[c][0.75cm][c]{1cm}{\centering NB}  & 0.66                                                                                    & 0.61                                                                                 & 0.58                                                                                        \\ \hline
			\multirow{5}{*}{CSE-CIC-IDS2018} & \parbox[c][0.75cm][c]{1cm}{\centering SVM} & 0.77                                                                                    & 0.69                                                                                 & 0.66                                                                                        \\ \cline{2-5} 
			& \parbox[c][0.75cm][c]{1cm}{\centering RF}  & 0.95                                                                                    & 0.95                                                                                 & 0.95                                                                                        \\ \cline{2-5} 
			& \parbox[c][0.75cm][c]{1cm}{\centering NB}  & 0.72                                                                                    & 0.61                                                                                 & 0.56                                                                                        \\ \hline
		\end{tabular}%
	}
	\label{table:Traditional evaluation metrics}
\end{table}

In the next step, we evaluate the traditional ML models based on detection with the enriched raw data (i.e., here, we do not use the pre-extracted features of datasets). The traditional ML models are evaluated over the pre-processed vectorized flows (see Table \ref{table:Traditional evaluation metrics vectorize}). Due to the dependence of the size of the SVM model to the size of the input data, and also according to the huge size of the pre-processed dataset, the SVM model cannot be fit in our server and it is impractical for the real world IDSs. Among the traditional ML models, the RF shows remarkable results than the others over the raw data.

\begin{table}[h]
	\centering
	\caption{The traditional ML models for the binary classification evaluated on the enriched raw data of CIC-IDS2017 and CSE-CIC-IDS2018.}
	\resizebox{\linewidth}{!}{%
		\begin{tabular}{|c|c|c|c|c|}
			\hline
			Dataset                          & \parbox[c][0.75cm][c]{1cm}{\centering Model}    & \parbox[c][0.75cm][c]{2cm}{\centering Precision} & \parbox[c][0.75cm][c]{2cm}{\centering Recall} & \parbox[c][0.75cm][c]{2cm}{\centering $F_{1}$ score} \\ \hline
			\multirow{3}{*}{CIC-IDS2017}     & \parbox[c][0.75cm][c]{1cm}{\centering RF}  & 0.96                                                                                    & 0.96                                                                                & 0.96                                                                                        \\ \cline{2-5} 
			& \parbox[c][0.75cm][c]{1cm}{\centering NB}  & 0.91                                                                                    & 0.91                                                                                 & 0.90                                                                                        \\ \hline
			\multirow{3}{*}{CSE-CIC-IDS2018} & \parbox[c][0.75cm][c]{1cm}{\centering RF}  & 0.90                                                                                    & 0.89                                                                                 & 0.89                                                                                        \\ \cline{2-5} 
			& \parbox[c][0.75cm][c]{1cm}{\centering NB}  & 0.75                                                                                    & 0.75                                                                                 & 0.75                                                                                        \\ \hline
		\end{tabular}%
	}
	\label{table:Traditional evaluation metrics vectorize}
\end{table}

The results of the multi-class RF model as the best traditional ML model over the enriched raw data are also presented in Table \ref{table:RF_Multi_category}.

\begin{table}[h]
	\centering
	\caption{The RF multi-class classifier evaluated on the enriched raw data of CIC-IDS2017 and CSE-CIC-IDS2018.}
	\resizebox{\linewidth}{!}{%
		\begin{tabular}{|c|c|c|c|c|}
			\hline
			Dataset                          & \parbox[c][0.75cm][c]{1cm}{\centering Category}    & \parbox[c][0.75cm][c]{2cm}{\centering Precision} & \parbox[c][0.75cm][c]{2cm}{\centering Recall} & \parbox[c][0.75cm][c]{2cm}{\centering $F_{1}$ score} \\ \hline
			\multirow{16}{*}{CIC-IDS2017}     & \parbox[c][0.75cm][c]{1cm}{\centering Benign}  & 0.97                                                                                    & 0.84                                                                                & 0.90                                                                                        \\ \cline{2-5} 
			& \parbox[c][0.75cm][c]{2cm}{\centering Botnet}  & 0.99                                                                                    & 0.94                                                                                 & 0.96   \\ \cline{2-5} 
			& \parbox[c][0.75cm][c]{2cm}{\centering Port Scan}  & 0.99                                                                                    & 0.99                                                                                 & 0.99   
			\\ \cline{2-5} 
			& \parbox[c][0.75cm][c]{2cm}{\centering DoS/DDoS}  & 0.96                                                                                    & 1.00                                                                                 & 0.98   
			\\ \cline{2-5} 
			& \parbox[c][0.75cm][c]{2cm}{\centering Heartbleed/ Infiltration}  & 0.00                                                                                    & 0.00                                                                                 & 0.00   \\ \cline{2-5} 
			& \parbox[c][0.75cm][c]{2cm}{\centering Brute Force}  & 0.98                                                                                    & 1.00                                                                                 & 0.99   \\ \cline{2-5} 
			& \parbox[c][0.75cm][c]{2cm}{\centering Web Attacks}  & 0.85                                                                                    & 0.91                                                                                 & 0.88
			\\ \cline{2-5} 
			& \parbox[c][0.75cm][c]{2cm}{\centering Total}  & 0.96                                                                                    & 0.96                                                                                 & 0.96                                                                                        \\ \hline
			\multirow{14}{*}{CSE-CIC-IDS2018} & \parbox[c][0.75cm][c]{2cm}{\centering Benign}  & 0.74                                                                                    & 0.93                                                                                 & 0.82                                                                                        \\ \cline{2-5} 
			& \parbox[c][0.75cm][c]{2cm}{\centering Heartbleed/ Infiltration}  & 0.33                                                                                    & 0.12                                                                                 & 0.18
			\\ \cline{2-5} 
			& \parbox[c][0.75cm][c]{2cm}{\centering Botnet}  & 0.99                                                                                    & 1.00                                                                                 & 0.99                                                                                        \\ \cline{2-5} 
			& \parbox[c][0.75cm][c]{2cm}{\centering DoS/DDoS}  & 0.95                                                                                    & 0.91                                                                                 & 0.93                                                                                        \\ \cline{2-5} 
			& \parbox[c][0.75cm][c]{2cm}{\centering Web Attacks}  & 1.00                                                                                    & 0.79                                                                                 & 0.88                                                                                        \\ \cline{2-5} 
			& \parbox[c][0.75cm][c]{2cm}{\centering Brute Force}  & 1.00                                                                                    & 0.93                                                                                 & 0.97
			\\ \cline{2-5} 
			& \parbox[c][0.75cm][c]{2cm}{\centering Total}  & 0.82                                                                                    & 0.84                                                                                 & 0.82                                                                                                                       \\ \hline
		\end{tabular}%
	}
	\label{table:RF_Multi_category}
\end{table}

\subsubsection{Determining the Hyper-Parameters for the DID-based LSTM}
As mentioned before, the sequential nature of flows, packets, and bytes leads us to use LSTM models for the DID framework. Hence, the next step is to determine the appropriate hyper-parameters of the proposed LSTM models.  The main hyper-parameters of an LSTM model are the number of layers and the number of units in each layer. The main results of the grid search to tune hyper-parameters are reported in the following. LSTM-1a has one LSTM layer with 50 units and some fully-connected layers with 2500, 1250, 512, 256, 64, and 16 neurons. LSTM-2a has the same fully-connected layers but two LSTM layers, whereas the number of units of the first LSTM layer is 100, and the second one is 50 units. LSTM-1b and LSTM-2b are similar to LSTM-1a and LSTM-1b respectively. The main difference between these two models comes from their fully connected layers. They have simpler fully-connected layers than ``a'' models (they only have two 64 and 16 neurons layers). In all models, all layers' activation functions (except the last one) are ReLU, and dropout layers with a 20\% drop rate are added among fully-connected ones.

\begin{table}[h]
	\centering
	\caption{The comparison of different LSTM models employed in the DID framework.}
	\resizebox{\linewidth}{!}{%
		\begin{tabular}{|c|c|c|c|c|} 
			\hline
			\parbox[c][][c]{2cm}{\centering Model}                             & \parbox[c][][c]{2cm}{\centering Precision}               & \parbox[c][][c]{2cm}{\centering Recall}                 & \parbox[c][][c]{2cm}{\centering $F_{1}$ score}           \\ 
			\hline\hline
			\parbox[c][0.5cm][c]{2cm}{\centering LSTM-1a}                                 & 0.989                  & 0.995                  & 0.991                   \\ 
			\hline
			\parbox[c][0.5cm][c]{2cm}{\centering LSTM-1b}                                 & 0.983                  & 0.990                  & 0.987                   \\ 
			\hline
			\parbox[c][0.5cm][c]{2cm}{\centering LSTM-2a}                                 & \textbf{0.992}                  &  \textbf{0.998}                  & \textbf{0.994}                   \\ 
			\hline
			\parbox[c][0.5cm][c]{2cm}{\centering LSTM-2b}                                 & 0.987                  & 0.993                  & 0.990                   \\ 
			\hline
		\end{tabular}
	}
	\label{table:DID evaluation metrics}
\end{table}

Table \ref{table:DID evaluation metrics} represents the results of different LSTM models introduced above, evaluated on the CIC-IDS2017 dataset. As the results show, LSTM-2a outperforms other models; however, the performance and simplicity of LSTM-1b can also be a good candidate for practical implementations.

\begin{figure}[h!]
	\centering
	\includegraphics[width=\linewidth, keepaspectratio]{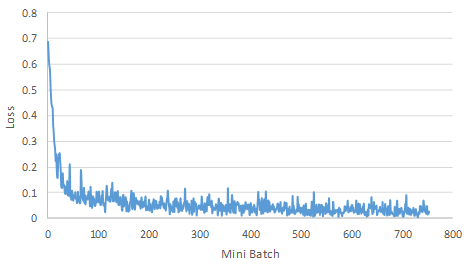}
	\caption{The loss of the proposed deep learning module (LSTM-2a) of DID during the training phase.}
	\label{fig:lstm-loss}
\end{figure}

Figure \ref{fig:lstm-loss} depicts the loss value in the training phase with selected hyper-parameters for LSTM-2a. At the end of the training phase, the mean of loss in training and validation data is $0.03$ and $0.01$, respectively. The lower value of loss in the validation phase is due to the dropout layers applied during the training phase, which improves the generalization of the deep model. Consequently, by removing them in the validation phase, better results are achieved. The results of the evaluation of this model by the test data are also presented in Table \ref{table:LSTM evaluation metrics} for comparison with previous works.

We also evaluated our model as a multi-class classifier. In this scenario, not only the type of traffic but also its category type will be determined. The performance of the multi-class classification over the attack categories of the CIC-IDS2017 and CSE-CIC-IDS2018 datasets are presented in Table \ref{table:DID_Multi_category}.

\begin{table}[h]
	\centering
	\caption{The LSTM based DID multi-class classifier evaluated on the enriched raw data of CIC-IDS2017 and CSE-CIC-IDS2018.}
	\resizebox{\linewidth}{!}{%
		\begin{tabular}{|c|c|c|c|c|}
			\hline
			Dataset                          & \parbox[c][0.75cm][c]{1cm}{\centering Category}    & \parbox[c][0.75cm][c]{2cm}{\centering Precision} & \parbox[c][0.75cm][c]{2cm}{\centering Recall} & \parbox[c][0.75cm][c]{2cm}{\centering $F_{1}$ score} \\ \hline
			\multirow{16}{*}{CIC-IDS2017}     & \parbox[c][0.75cm][c]{1cm}{\centering Benign}  & 0.99                                                                                    & 0.94                                                                                & 0.97                                                                                        \\ \cline{2-5} 
			& \parbox[c][0.75cm][c]{2cm}{\centering Botnet}  & 0.86                                                                                    & 1.00                                                                                 & 0.92   \\ \cline{2-5} 
			& \parbox[c][0.75cm][c]{2cm}{\centering Port Scan}  & 0.99                                                                                    & 0.99                                                                                 & 0.99   
			\\ \cline{2-5} 
			& \parbox[c][0.75cm][c]{2cm}{\centering DoS/DDoS}  & 1.00                                                                                    & 1.00                                                                                 & 1.00   
			\\ \cline{2-5} 
			& \parbox[c][0.75cm][c]{2cm}{\centering Heartbleed/ Infiltration}  & 0.00                                                                                    & 0.00                                                                                 & 0.00   \\ \cline{2-5} 
			& \parbox[c][0.75cm][c]{2cm}{\centering Brute Force}  & 1.00                                                                                    & 1.00                                                                                 & 1.00   \\ \cline{2-5} 
			& \parbox[c][0.75cm][c]{2cm}{\centering Web Attacks}  & 0.95                                                                                    & 0.98                                                                                 & 0.96
			\\ \cline{2-5} 
			& \parbox[c][0.75cm][c]{2cm}{\centering Total}  & 0.99                                                                                    & 0.99                                                                                 & 0.99                                                                                        \\ \hline
			\multirow{14}{*}{CSE-CIC-IDS2018} & \parbox[c][0.75cm][c]{2cm}{\centering Benign}  & 0.96                                                                                    & 0.77                                                                                 & 0.85                                                                                        \\ \cline{2-5} 
			& \parbox[c][0.75cm][c]{2cm}{\centering Heartbleed/ Infiltration}  & 0.53                                                                                    & 0.90                                                                                 & 0.67
			\\ \cline{2-5} 
			& \parbox[c][0.75cm][c]{2cm}{\centering Botnet}  & 1.00                                                                                    & 1.00                                                                                 & 1.00                                                                                        \\ \cline{2-5} 
			& \parbox[c][0.75cm][c]{2cm}{\centering DoS/DDoS}  & 0.99                                                                                    & 1.00                                                                                 & 1.00                                                                                        \\ \cline{2-5} 
			& \parbox[c][0.75cm][c]{2cm}{\centering Web Attacks}  & 0.95                                                                                    & 0.97                                                                                 & 0.96                                                                                        \\ \cline{2-5} 
			& \parbox[c][0.75cm][c]{2cm}{\centering Brute Force}  & 1.00                                                                                    & 1.00                                                                                 & 1.00
			\\ \cline{2-5} 
			& \parbox[c][0.75cm][c]{2cm}{\centering Total}  & 0.93                                                                                    & 0.90                                                                                 & 0.90                                                                                                                      \\ \hline
		\end{tabular}%
	}
	\label{table:DID_Multi_category}
\end{table}

\begin{table}[h]
	\centering
	\caption{The results achieved by the proposed DID framework on the CIC-IDS2017 and CSE-CIC-IDS2018 alongside its comparison to \cite{zavrak2020anomaly}, McPAD, and Deepcoin which are evaluated on the CIC-IDS2017. Besides, the reported values of \cite{Soheily2017ISCX}, \cite{kakavand2016effective}, \cite{khan2019scalable}, and  \cite{zeng2019deep} over ISCX IDS 2012 are presented.}
	\resizebox{\linewidth}{!}{%
		\begin{tabular}{|c|c|c|c|c|c|} 
			\hline
			\parbox[c][0.5cm][c]{1.5cm}{\centering Dataset}                         &  \parbox[c][0.5cm][c]{2cm}{\centering Model}                             & \parbox[c][0.5cm][c]{1.3cm}{\centering Precision}              & \parbox[c][0.5cm][c]{1.3cm}{\centering Fall-out}               & \parbox[c][0.5cm][c]{0.9cm}{\centering Recall}                 & \parbox[c][0.5cm][c]{1.3cm}{\centering $F_{1}$ score}           \\ 
			\hline\hline
			\multirow{2}{*}{ \parbox[c][][c]{1.5cm}{\centering CSE-CIC \\IDS 2018}} & \parbox[c][0.5cm][c]{4cm}{\centering \textbf{DID} (with LSTM)}                   & \textbf{0.933}                  & \textbf{0.105}                  & \textbf{0.923}                  & \textbf{0.927}                   \\
			\cline{2-6}
			& \parbox[c][0.5cm][c]{4cm}{\centering Random Forest}                      & 0.902                  & 0.196                  & 0.891                  & 0.892      \\
			\hline
			\multirow{5}{*}{ \parbox[c][][c]{1.5cm}{\centering CIC-IDS \\2017}} & \parbox[c][0.5cm][c]{4cm}{\centering \textbf{DID} (with LSTM)}                   & 0.992                  & \textbf{0.002}                  & \textbf{0.998}                  & \textbf{0.994}                   \\ 
			\cline{2-6}
			& \parbox[c][0.5cm][c]{4cm}{\centering Zavrak et al. \cite{zavrak2020anomaly}}                      & -                  &  0.550                 & 0.950                  & -                   \\ 
			\cline{2-6}
			& \parbox[c][0.5cm][c]{4cm}{\centering McPAD \cite{Perdisci2009McPAD}}                      & \textbf{0.993}                  & 0.019                  & 0.177                  & 0.300                   \\ 
			\cline{2-6}
			& \parbox[c][0.5cm][c]{4cm}{\centering DeepCoin \cite{ferrag2019deepcoin}}                      & 0.983                  & 0.009                 & -                  & -                   \\
			\hline
			\multirow{5}{*}{\parbox[c][0.75cm][c]{1.5cm}{\centering ISCX IDS\\ 2012}}                   & \multicolumn{1}{c|}{\parbox[c][0.5cm][c]{5cm}{\centering Soheily-Khah and et al. \cite{Soheily2017ISCX}}} & \multicolumn{1}{c|}{0.987} & \multicolumn{1}{c|}{\textbf{0.001}} & \multicolumn{1}{c|}{0.989} & \multicolumn{1}{c|}{\textbf{0.988}}  \\
			\cline{2-6}
			& \parbox[c][0.5cm][c]{4cm}{\centering PCA-based TMAD \cite{kakavand2016effective}}                      & \textbf{0.999}                  & 0.012                  & 0.970                  & 0.984                   \\ 
			\cline{2-6}
			& \parbox[c][0.5cm][c]{4cm}{\centering Kahn and et al. \cite{khan2019scalable}}                      & 0.972                  & 0.007                  & 0.975                  & 0.973                   \\
			\cline{2-6}
			& \parbox[c][0.5cm][c]{4cm}{\centering DFR \cite{zeng2019deep}}                      & 0.981                  & -                  &\textbf{0.991}                  & 0.986                   \\
			\hline
		\end{tabular}
	}
	\label{table:LSTM evaluation metrics}
\end{table}

\subsubsection{Comparison with Similar Researches}
Our work is comparable with two categories of related works. The first category belongs to studies evaluated on ISCX/CIC IDS datasets and the second one concerns those that focus on the contents of the traffic payloads. Soheily-Khah \etal~\cite{Soheily2017ISCX} use $50$ features of the ISCX 2012 dataset to evaluate their model, which is achieved by combining $K$-means and random forest algorithms. This research is somehow comparable to our work since it uses some learning algorithms over the ISCX dataset, and this method can be compared with deep learning. Their model has achieved recall and fall-out of around $98.9$ and $0.1$, respectively. Note that since they have not announced the average evaluation metrics, we have used their reported tables and the mean of their metrics for different protocols (since PR was not reported, we calculate its value). The first category also comprises some other works such as \cite{khan2019scalable}, \cite{zeng2019deep}, and \cite{ferrag2019deepcoin}, which all focus on using deep-learning methods over features that are extracted from ISCX/CIC IDS datasets. The main point that makes related works on ISCX IDS 2012 comparable with studies on CIC-IDS2017 is that the CIC 2017 just an updated version of ISCX 2012. It includes more benign profiles and attack version \cite{Sharafaldin2018ISCX}. Due to the more complexity and completeness of the CIC-IDS2017, the produced results are more reliable.

In the second category, we have works like \cite{Perdisci2009McPAD}, which is one of the best research studies about detecting content-based attacks. We have evaluated this method by using its source code, which is available at \cite{mcpadcode}. Even though this code yields good results over the dataset used by themselves in \cite{Perdisci2009McPAD}, but it shows a weak performance in learning CIC-IDS2017 dataset. This weakness is related to the ISCX/CIC IDS dataset's comprehensiveness against previous ones like DARPA or KDD99. The results of the evaluation of McPAD by CIC-IDS2017 are also represented in Table \ref{table:LSTM evaluation metrics}.

The main weakness of McPAD is its detection rate, represented by the recall, which is around $20\%$. Although McPAD has a significant detection rate over the dataset used in \cite{Perdisci2009McPAD}, it cannot be beneficial in real-world traffic. Further inspections show that while their benign traffic is suitable, the attack ones used for the evaluation have some notable weaknesses. For example, allShellcode.pcap file has only 11 TCP sessions, which in each one contains a shell-code attack. As the NOP sled in these attacks has many repetitions of bytes like 0x90 and 0x61, they can be easily detected. Besides, the other attack file, which is called allGeneric.pcap, has 66 HTTP attacks. Among them, 11 shell-code attacks can be detected as the previous one, and the others have hostnames that do not exist in the training dataset (like www and www.i-pi.com). Consequently, the $n$-gram mechanism can detect these kinds of attacks. However, in the case of the CIC-IDS2017 dataset, although its alarm has significant reliability ($\mathrm{PR}=99.3\%$), its detection rate is low ($\mathrm{RC}=17.7\%$).

Another related work in the second category is \cite{kakavand2016effective}, which focuses on extracting features of HTTP packet payloads by the PCA algorithm. Finally, using a Text Mining-based Anomaly Detection (TMAD) model tries to detect attack traffics.

Zavrak et al. \cite{zavrak2020anomaly} use variational autoencoders as anomaly-based IDS sensors. Their evaluation has been done over the CIC-IDS2017 dataset. The anomaly model is trained by the records from Monday's CSV file that only contains benign traffic. Other days of the dataset are used for the test phase. Finally, the FPR and TPR metrics are reported that are equivalent to the FO and RC, respectively.

Finally, we would like to report the resource and time consumption of the proposed model and the comparison with the RF model as the best traditional ML model over the enriched raw network data. The RF model requires much lower resources in comparison with the LSTM model according to our evaluations. On average, the RF model needs about 208MB memory and 100\% usage of one CPU core of the machine. In contrast, the LSTM model uses about 900MB memory and 20\% of the GPU processing power for the same training data records. As presented in Table \ref{tab:Timing}, the RF model is much faster in the training phase than the LSTM model. But the two models are competitive in the test phase. Since an IDS is mainly used in the test mode in the real world, the training time can have less weight on the final comparison of these two models. On the other side, as shown in Tables \ref{table:RF_Multi_category} and \ref{table:DID_Multi_category}, the strength of the LSTM model can be highlighted in more complicated dataset like CSE-CIC-IDS2018. Definitely, in the real world, there are complicated data that makes the LSTM model more applicable. However, for a more relaxed scenario where the RF model has an acceptable performance and requires lower hardware resources, the RF might be a better choice for the IDS implementation.  

\begin{table}
	\caption{The average processing time per flow in the training and test phases.} \label{tab:Timing}
	\centering
	\resizebox{\linewidth/3*2}{!}{%
		\begin{tabular}{|c|c|c|c|}
			\hline
			Model                & \parbox[c][0.5cm][c]{1cm}{Type}        & Training (sec) & Test (sec) \\ \hline
			\multirow{3}{*}{DID} & \parbox[c][0.5cm][c]{1cm}{Binary}      &    0.014      &   0.007   \\ \cline{2-4} 
			& \parbox[c][0.5cm][c]{1.75cm}{Multi-Class} &    0.016      &   0.008   \\ \hline
			\multirow{3}{*}{RF}  & \parbox[c][0.5cm][c]{1cm}{Binary}      &     0.004     &   0.004   \\ \cline{2-4} 
			& \parbox[c][0.5cm][c]{1.75cm}{Multi-Class} &     0.005     &   0.005   \\ \hline
		\end{tabular}
	}
\end{table}

The input data to DID are flows and time consumption of evaluation of each flow is around 7 milliseconds. According to \cite{Jurkiewicz2018flowdist}, on average, we can assume each flow contains 78 packets, and each packet contains 870 bytes. As a result, the proposed model in our test environment can handle around $75$ Megabit per second traffic data per GPU. This can be a challenge in high-performance applications. However, by applying various optimizations, the above performance can be significantly increased and this can be an interesting direction for future works.

\subsection{Discussion}
According to our experiments, the proposed deep intrusion detection (DID) approach can have a comparative advantage over previous works in inspecting more varieties of attacks, especially those who manipulate the payload of traffic. However, the proposed approach has some challenges which should be addressed in future works. Some of these challenges are discussed below.

The main current shortcoming of using deep learning in network detection is its throughput. By increasing the Internet bandwidth, we have to count on devices with high throughput along with a high detection rate and low false alarm. Consequently, according to the complexity of deep learning algorithms, one of the main forward steps toward this goal is to optimize the deep intrusion detectors and implement them over high-performance devices like FPGAs or ASICs.

One of the most challenging issues in the scope of intrusion detection systems is analyzing the encrypted traffics. Since the content of encrypted flows is randomized, most of the signature-based IDSes have significant issues with these kinds of traffics. Evaluation of DID framework over encrypted traffics can be studied in future works.

Another challenge to making ML-based IDSes more applicable in practice is to adapt them to imbalanced data. The imbalance of data can make a machine learning model tend to the ``more observed'' (major) category. However, detecting the minor category may be of high value for us (such as detecting cancer in medical applications or attack detection in computer networks). Alongside, if the test dataset is also imbalanced, the overall detection rate of the algorithm cannot provide a useful measure of the performance of the intrusion detection method in real scenarios. For example, for a dataset with 95\% benign traffic, this can lead to a model that labels all the inputs as benign traffic to achieve $95\%$ accuracy while the desired goal of the intrusion detection system is to detect attacks as much as possible with low false positive. In this paper, the data reduction mechanism for the majority group has been applied. However, this solution can cause some losses in the diversity of the major category (\ie, in this paper, the benign traffic). Consequently, some kinds of benign flows may be detected as attacks in a more comprehensive dataset, which has more complicated attacks. 

Finally, in this research, we have used a labeled dataset for training the model. However, the lack of adequate diversity in this dataset can lead to poor performance in real networks. On the other hand, each network has its own behavior for normal traffics (like the number of new connections per second), which may be considered an abnormal behavior in other networks. Hence, it is very crucial that we learn the models according to their deployment environment.

\section {Conclusion} \label{sec:conclusion}
This paper presented a Deep Intrusion Detection approach that uses deep learning algorithms for detecting a wide range of attacks, including content-based ones like SQL injection and Heartbleed attack. We have used an LSTM-based model as an implementation of the deep learning module of the DID approach. LSTM layers can extract meaningful relations among bytes of packets of each flow. Besides using dropout layers, we tried to avoid overfitting. Four metrics that provide valuable information in intrusion detection applications have been selected for evaluation, namely, precision, recall, fall-out, and $F_{1}$ score. On the CIC-IDS2017 dataset, we have achieved a precision of $0.992$, fall-out of $0.2$, recall of $0.998$, and $F_{1}$ score of $0.994$. Furthermore, on the CSE-CIC-IDS2018, the recall of $0.923$ and precision of $0.933$ achieved. The experimental results show that the proposed approach has better performance than the previous works.

\begin{acknowledgements}
The authors would like to thank Ramin Shirali and Jafar Gholamzadeh for their invaluable help, discussion, and feedback on this work.
\end{acknowledgements}

\section*{Compliance with Ethical Standards}
\textbf{Funding:} No funding was received to assist with the preparation of this manuscript.\\
\textbf{Conflicts of interests:} The authors have no conflicts of interest to declare that are relevant to the content of this article.\\
\textbf{Ethical approval:} This article does not contain any studies with human participants or animals	performed by any of the authors.

\bibliographystyle{spmpsci}      
\bibliography{references}   

%
%

\end{document}